\documentclass[ aip,
 jmp,%
 amsmath,amssymb,
 reprint,%
]{revtex4-1}

\usepackage{graphicx}

\usepackage{stmaryrd}
\usepackage{amsmath}	 
\usepackage{amsfonts} 
\begin{document}


\title{Trampoline metamaterial: Local resonance enhancement by springboards
} 



\author{Osama R. Bilal}
\noaffiliation
\author{Mahmoud I. Hussein}
\email[]{mih@colorado.edu}
\noaffiliation
\affiliation{Department of Aerospace Engineering Sciences, University of Colorado Boulder, Boulder, Colorado 80309, USA }

\date{\today}

\begin{abstract}
We investigate the dispersion characteristics of locally resonant elastic metamaterials formed by the erection of pillars on the solid regions in a plate patterned by a periodic array of holes. We show that these solid regions effectively act as springboards leading to an enhanced resonance behavior by the pillars when compared to the nominal case of pillars with no holes. This local resonance amplification phenomenon, which we define as the \textit{trampoline effect}, is shown to cause subwavelength band gaps to increase in size by up to a factor of 4. This outcome facilitates the utilization of subwavelength metamaterial properties over exceedingly broad frequency ranges.   
\end{abstract}

\pacs{}

\maketitle 

Phononic crystals and locally resonant acoustic/elastic metamaterials have been the focus of extensive research efforts in recent years due to their attractive dynamical characteristics, such as the possibility of exhibiting band gaps. In a phononic crystal, band gaps are generated by Bragg scattering for which an underlying constraint is that the wavelength has to be on the order of the lattice spacing. \cite{kushwaha1993acoustic, sigalas1993band} In a locally resonant acoustic/elastic metamaterial, on the other hand, band gaps may be generated by the mechanism of hybridization between local resonances and the  dispersion properties of the underlying medium, and this in turn may take place at the subwavelength regime. \cite{liu2000locally} \\
\indent While in principal periodicity is not a necessity in a metamaterial, the introduction of the locally resonant elements in a symmetric fashion enables intrinsic, unit-cell based, description of the wave propagation characteristics, in addition to attaining the benefits of order and compactness. The presence of periodicity, in itself, produces direction-dependent frequency bands and band gaps (caused by Bragg scattering). These unique dispersion properties may sufficiently be utilized in numerous applications involving wave filtering,~\cite{pennec2004tunable,HusseinIMECE2005,Hussein_SMO_2006} localization,~\cite{sigalas1997elastic,Torres_1999} guiding,~\cite{Torres_1999,khelif2004guiding} focusing,~\cite{Cervera2001,Zhang_2004,Yang_2004} collimation,~\cite{Chen_2004,Christensen_2006} among others. The added feature of local resonance, however, gives rise to a qualitatively different type of dynamical behavior, such as negative effective elastic moduli and/or density,~\cite{Li:2004sr,Fang:2006mw,Avila_MMS_2008,Ding:2007kk} along with the possibility of generation of subwavelength band gaps.\cite{liu2000locally} The applications of locally resonant acoustic/elastic metamaterials are, in turn, far from conventional, e.g., subfrequency wave isolation,~\cite{Ho_APL_2003} subwavelength focusing and imaging,~\cite{guenneau2007acoustic,ambati2007surface} and cloaking,~\cite{torrent2008acoustic} to name a few. Crossing the boundaries of acoustics and elasticity, our group at CU-Boulder has recently proposed the utilization of locally resonant metamaterials for the control of heat in a semiconducting thin-film.~\cite{Davis_Hussein_2013} Regardless of the application, one of the generally desirable characteristics is for the metamaterial to exhibit a large band gap. While the problem of unit-cell optimization for maximum band-gap size has been actively pursued for phononic crystals,~\cite{sigmund2003systematic,Hussein_SMO_2006,Bilal_PRE_2011} less work has been done in exploring new approaches for band-gap enlargement in locally resonant acoustic/elastic metamaterials.\\ 
\indent One of the practical realizations of locally resonant elastic metamaterials is the construction of a periodic array of pillars on a plate \cite{pennec2008low, wu2008evidence}. In this configuration, the pillars by virtue of their dynamic stiffness serve as the local resonators as they are essentially rod/beam-like structures laid out on a flexural foundation. The diameter and height of the pillars determine their resonant frequencies, and the extent of the coupling of the pillar modes with the foundational plate modes is dependent primarily on the plate stiffness and thickness as well as the lattice spacing of the pillars.
\begin{figure*}[ht]
\centering
\includegraphics[scale=0.95]{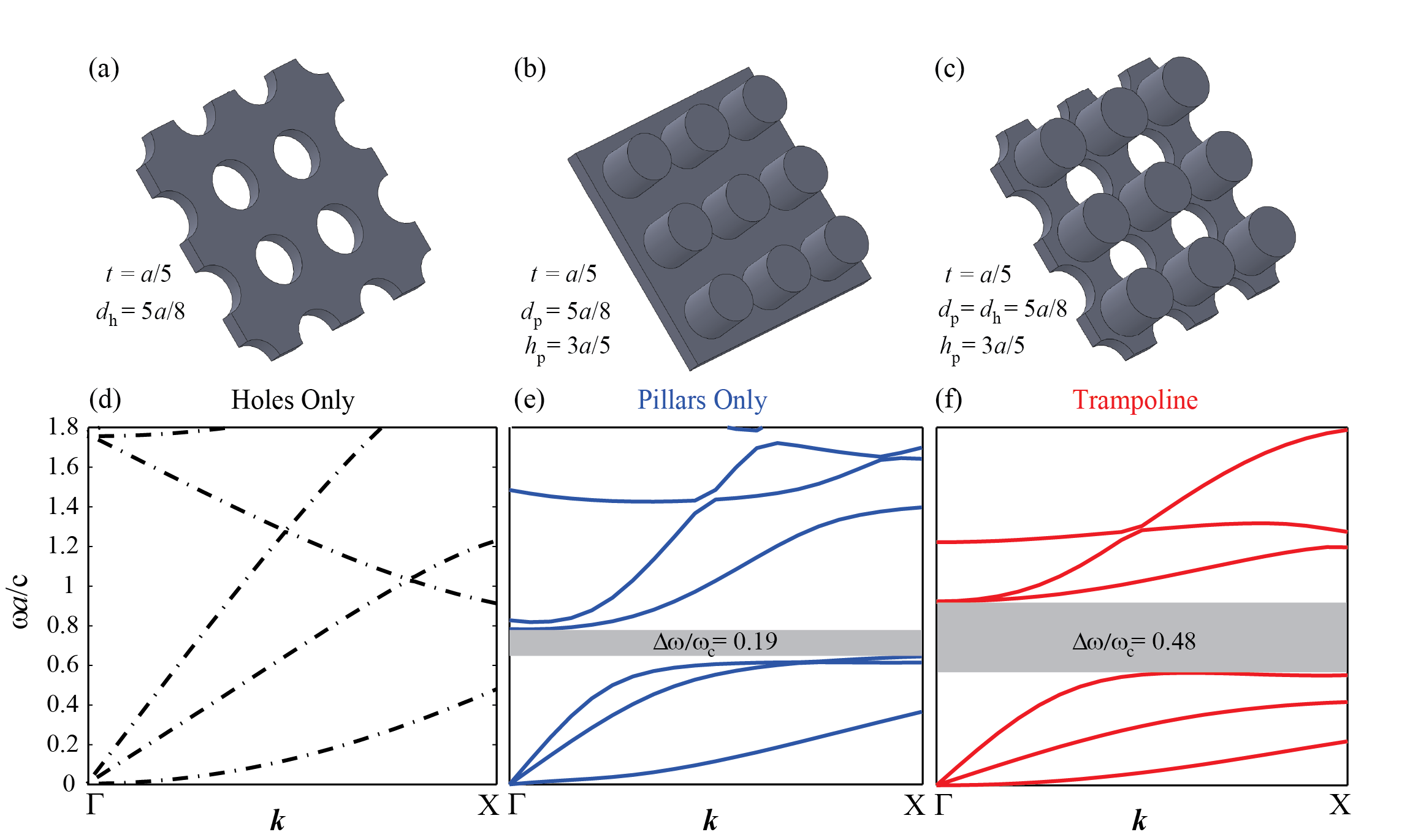}
\caption{Illustration of the concept of a trampoline metamaterial. The top row displays schematics of (a) a standard phononic crystal (consisting of a periodic array of holes in a plate), (b) a standard pillared elastic metamaterial (consisting of a periodic array of pillars on a plate), and (c) a trampoline metamaterial (consisting of a periodic array of pillars on a plate intertwined with a periodic array of holes). The frequency band structure of the three configurations in the $\Gamma \rm X$ direction is shown in the bottom row. In the frequency range displayed, the phononic crystal has no band gaps (d), the standard pillared metamaterial exhibits a subwavelength band gap with a relative size of 0.19 (e), and the proposed trampoline metamaterial exhibits an enhanced subwavelength band gap with a relative size of 0.48.}
\label{fig:01}
\end{figure*}

Driven by the desire for band-gap amplification, in this work we propose a configuration formed by merging a periodically pillared structure, based on a plate foundation (Fig. \ref{fig:01}b), with a standard phononic crystal formed by removal of a periodic array of holes in a plate (Fig. \ref{fig:01}a). The emerging configuration is a locally resonant elastic metamaterial consisting of pillars erected on the solid regions of a phononic crystal plate (Fig. \ref{fig:01}c). In this configuration, each pillar is now rooted in a more compliant base due to the presence of the holes. This base effectively acts as a \textit{springboard} that allows the pillars' resonant motion to be enhanced. Due to the analogy to a recreational trampoline, we refer to this hybrid configuration as a \textit{trampoline metamaterial} and to the underlying mechanism as the \textit{trampoline effect}.

\begin{figure}[ht]
\centering
\includegraphics[scale= 1]{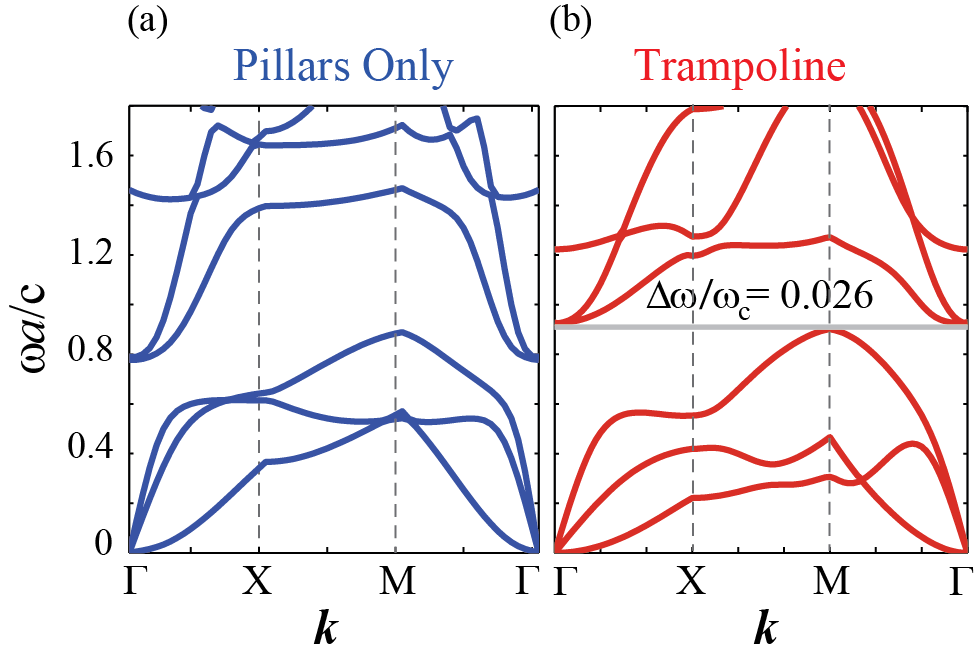}
\caption{Band structure along the entire $\Gamma \rm X \rm M \Gamma$ path of the irreducible Brillouin zone for cases shown in Figs. \ref{fig:01}b and \ref{fig:01}c. The standard pillared elastic metamaterial has no complete band gap, while the trampoline metamaterials has a band gap with a relative size of  0.026.}
\label{fig:02}
\end{figure}

To examine the behavior of the proposed trampoline metamaterial, we investigate the band structure characteristics for the propagation of Lamb waves in the three types of periodic materials shown in Figs. \ref{fig:01}a-\ref{fig:01}c. We choose silicon as the constitutive material ($\rho$ = 2330 $Kg/m^{3}$, $\lambda$ = 85.502 GPa, $\mu$ = 72.835 GPa) and for all configurations consider a plate thickness of $t$ = $a/5$ where $a$ is the lattice spacing. We set both the hole and pillar diameters as  $d_h = d_p = 5a/8$, and the pillar height as $h_p = 3a/5$. Using the theory of elasticity, we set up the equations of motion for all three models in the form $\nabla \cdot \mathbf{C}:\nabla^{\rm{S}}\mathbf{u}=\rho \ddot{\mathbf{u}}$ where $\mathbf{u}$ is the displacement vector, $\mathbf{C}$ is the elasticity tensor, $\rho$ is the density, $\nabla^{\rm{S}}$ is the symmetric gradient operator defined as $\nabla^{\rm{S}}=\left(\nabla\mathbf{u}+\nabla(\mathbf{u})^{\rm{T}}\right)/2$ and the dot is the inner product symbol. Denoting the position vector by $\mathbf{x}$, the wavevector by $\mathbf{k}$, the frequency by $\omega$ and the time by $t$, we apply Bloch's theorem, $\mathbf{\mathbf{u}}\mathbf{(x,k};t) = \mathbf{\mathbf{u}}\mathbf{(x,k)}e^{i(\mathbf{k}^T\mathbf{x}-\omega t)}$, to the equations of motion and obtain~\cite{Hussein_2009}
\begin{equation}
 \nabla . \mathbf{C} :\left[\nabla^{S}\mathbf{\tilde{u}} + \frac{i}{2}(\mathbf{k}^T \varotimes \mathbf{\tilde{u}} + \mathbf{k} \varotimes \mathbf{\tilde{u}}^T)\right]= -\rho\omega^{2}\mathbf{\tilde{u}},
\label{EVP}
\end{equation}
\noindent where $\mathbf{\tilde{u}}$ is the Bloch displacement vector and the symbol $\varotimes$ denotes the outer product. With the application of periodic boundary conditions on $\mathbf{\tilde{u}}$, Eq.~\eqref{EVP} forms the Bloch eigenvalue problem. We solve this problem using the finite element method  and obtain the band structure for each of the three configurations, which we show also in Fig. \ref{fig:01} for the standard phononic crystal (Fig. \ref{fig:01}d), the standard pillared elastic metamaterial (Fig. \ref{fig:01}e) and the proposed trampoline metamaterial (Fig. \ref{fig:01}f). In all our dispersion diagrams, the frequency is expressed as $\omega a/c$ where $c=\sqrt{E/\rho}$. We observe that the relative size of the locally resonant subwavelength band gap (along the $\Gamma \rm X$ direction) gets enlarged in size by a factor of approximately 2.5 when we compare the results of the trampoline metamaterial to the standard pillared elastic metamaterial. When examining the presence of a complete band gap along the $\Gamma \rm X \rm M \Gamma$ path of the irreducible Brillouin zone, we find that the standard pillared elastic metamaterial has no band gap, whereas the trampoline metamaterial exhibits a band gap with a relative size of 0.026 (see Fig. \ref{fig:02}). This outcome, for both the partial and complete band gaps, is due to the increased compliance of the pillar foundation (i.e., the pillar's plate base) due to the presence of the holes. The holes cut through the elastic foundation, reducing its area, and consequently allow not only the main body of the pillars to resonate but also portions of the foundation\textemdash or root\textemdash that each pillar is attached to. As such, the segments of the plate material between the holes effectively act as springboards that allow each pillar to resonate more intensely as they themselves contribute to the resonant motion.\\  
\begin{figure}[ht]
\centering
\includegraphics[scale= .82]{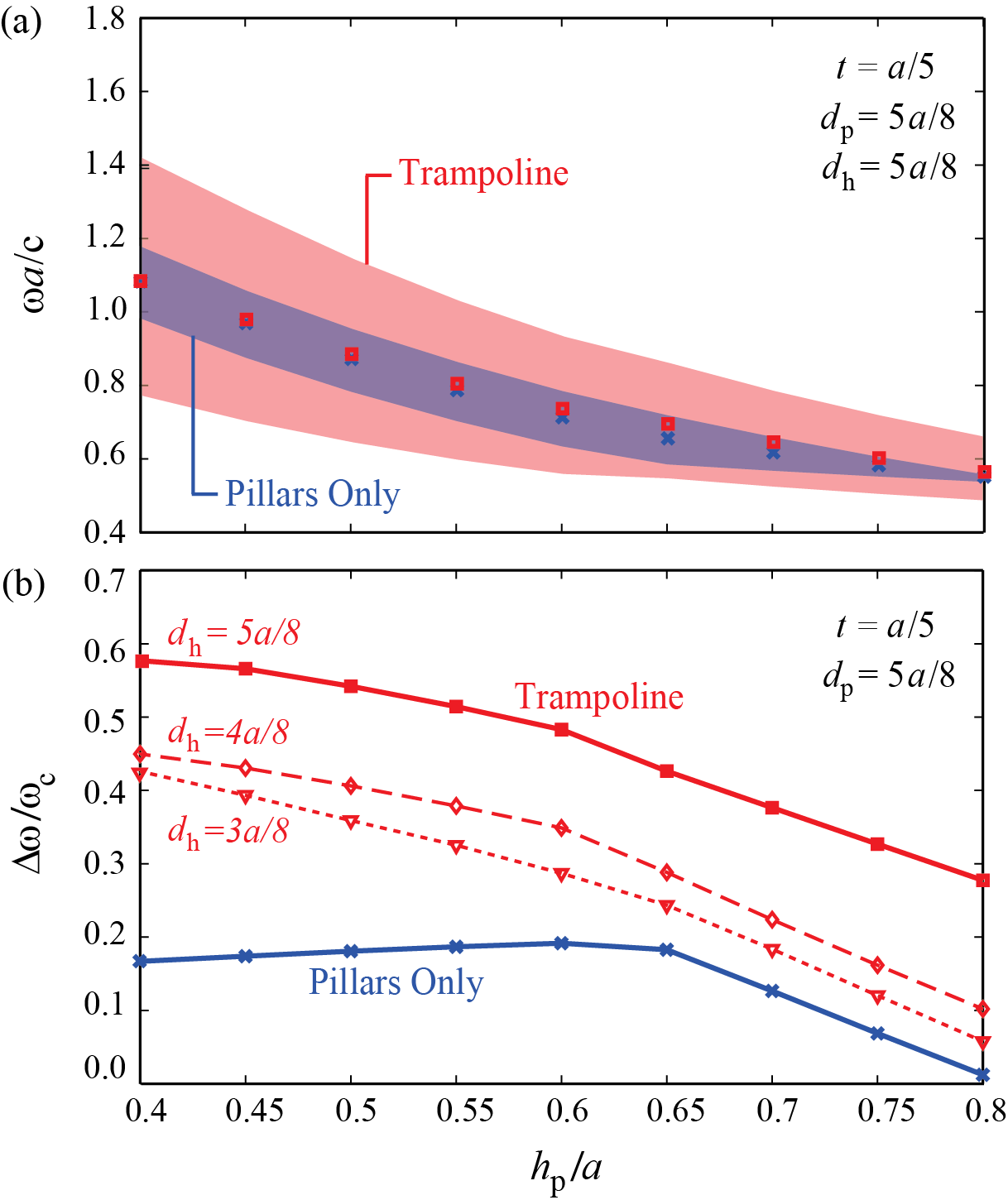}
\caption{Map of absolute (a) and relative (b) band gap as a function of normalized pillar height for a trampoline metamaterial compared to a standard pillared metamaterial. The trampoline effect results in a magnification of the subwavelength locally resonant band gap by a factor ranging from roughly 2 to 4 for geometries where the original $\Delta{\omega}/\omega_c$ is greater or equal to 0.1.}
\label{fig:03}
\end{figure}
\indent To further examine the trampoline effect, we analyze the problems shown in Figs. \ref{fig:01}b and \ref{fig:01}c but now for a range of values of pillar height and hole diameter. Figure \ref{fig:03}a presents a map of the size and location of the lowest band gap (also along the $\Gamma \rm X$ direction) as a function of pillar height for $d_h=5a/8$. The same results are again presented in Fig. \ref{fig:03}b but in the form of a plot of relative band-gap size, $\Delta{\omega}/\omega_c$, versus pillar height, where $\omega_c$ is the band-gap central frequency. We focus our attention to $d_p/a \geq 0.4$ because at lower pillar heights the lowest resonance frequency is high enough to interact with the Bragg scattering regime (which is outside the scope of this Letter). The blue solid line in Fig. \ref{fig:03}b is for the standard pillared elastic metamaterial. We observe that as the pillar height for this ``pillars only" case increases, the relative band-gap size also increases but only up to the point where $h_p/a=0.65$, after which it descends in value. This non-monatomic behavior is due to the coupling of the pillar vibrations with the flexural motion of the base plate, which becomes more profound as $h_p/a$ increases. The dotted, dashed and solid red lines represent the relative band-gap size for the trampoline metamaterial configuration for different values of $d_h$. In addition to $d_h=5a/8$, two other hole diameters are considered, namely, $d_h=3a/8$ and $d_h=4a/8$ (noting a constant increment of $a/8$ between the three diameters). We observe that for the two additional trampoline metamaterial cases, the relative band-gap size is larger than the case of ``pillars only". This shows that the trampoline effect is present also in the lower hole diameter cases. It is noteworthy, however, that the incremental enlargement of relative band-gap size is more significant at higher values of hole diameter, which is explained by the quadratic decrease in springboard area as $d_h$ is increased. We also observe that all three cases display a monotonic trend suggesting that the coupling between the pillars and the base plate is significant across the entire range of $h_p/a$ considered, which is a characteristic of the trampoline effect. At $h_p/a$ values less than 0.4, the relative band-gap size for the trampoline metamaterials decreases, with this backward descend initiating at different points depending on $d_h$.\\ 
\indent Most noticeable in Fig. \ref{fig:03} is the remarkable values of relative band gap size that are attainable by the trampoline metamaterials. For the parameter set of $d_h=5a/8$ and  $h_p/a=0.4$, the size of the relative band gap is close to 0.6 (i.e., $60\%$). This is a significant improvement over reported relative band gap sizes in the literature for acoustic/elastic metamaterials in general and specifically for pillared metamaterials, which usually fall within the range of $20\%-30\%$ for a partial band gap.~\cite{pennec2008low, wu2008evidence} Upon analysing the entire $\Gamma \rm X \rm M \Gamma$ wavevector path, we note that for all the $h_p/a$ values considered in Fig. \ref{fig:03} there is a complete band gap for the standard pillared metamaterial only within the range of $0.40 \leq h_p/a < 0.50$, with a maximum relative size of 0.11 at $h_p/a=0.40$. For the trampoline metamaterial, on the other hand, a complete band gap exists within the range $0.40 \leq h_p/a < 0.62$ with a maximum relative size of 0.17 at $h_p/a=0.40$. Thus the trampoline effect both increases the size of the complete band gap and extends the $h_p/a$ range of its existence$-$an outcome that is favorable for many subwavelength applications that require wave attenuation both at a broad frequency range and along all spatial directions.\\
\begin{figure}[t!]
\centering
\includegraphics[scale= 1]{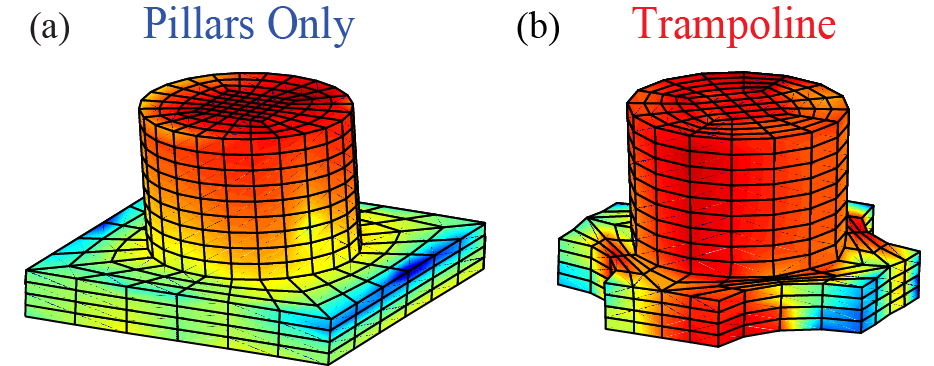}
\caption{Bloch mode shapes at $k_{x}= 0.62\pi/a$ for (a) standard pillared metamaterial and (b) trampoline metamaterial. The colormap represents the spatial distribution of the von Mises stress (in log scale).}
\label{fig:04}
\end{figure} 
\indent The trampoline effect can be further examined by viewing the Bloch mode shapes. A thorough analysis on the changes to the Bloch mode shapes when pillars are added to a uniform plate has been done by Wu and coworkers~\cite{wu2008evidence}. Here we focus only on the effects of adding holes to an already pillared plate. Figure \ref{fig:04} displays Bloch mode shapes for the standard ``pillars only" case (Fig. \ref{fig:04}a) and the trampoline case (Fig. \ref{fig:04}b) at $k_{x}= 0.62\pi/a$ (where $k_{x}$ denotes the wavenumber along the $\Gamma \rm X$ direction) for the third acoustic branch shown in Figs. \ref{fig:01}e and \ref{fig:01}f. In the figure, the total displacement is normalized by its own maximum value and the colormap represents the spatial distribution of the von Mises stress (in log scale). It can be noticed that in the case of the standard pillared elastic metamaterial (Fig. \ref{fig:04}a), the stress is concentrated within the main body of the pillar with very moderate stress levels in the base plate. This implies that there is little motion in the pillar foundation, and that the resonant motion is taking place mostly in the main body of the pillar. On the other hand, in the case of the trampoline metamaterial (Fig. \ref{fig:04}b), the stress distribution is high within both the main body of the pillar and in its foundation. This contrast in response confirms that the introduction of the holes indeed allows the plate base to act as a springboard, and this induces the trampoline effect and causes the band-gap enlargement observed in Fig. \ref{fig:01}f and in Fig. \ref{fig:03}.\\
\indent The concept of a trampoline metamaterial provides promising opportunities for the many applications that require large band gaps, and the proposed configuration can be easily fabricated and characterized considering that experimental investigations have already been conducted on plates consisting of either holes,~\cite{Zhang_APL_2006} or pillars.~\cite{Wu_PRB_2009,badreddine2012broadband} While the focus here has been on pillared elastic metamaterials and subwavelength band gaps, the concept is in principle applicable to other configurations (e.g., replacing the pillars with heavy inclusions) and to superwavelength band gaps. For higher frequency band gaps, a mixing between the hybridization and Bragg scattering mechanisms could lead to even larger amplifications in band-gap size. Furthermore, use of double pillars,~\cite{badreddine2012enlargement} multi-material pillars,~\cite{badreddine2012enlargement,badreddine2012broadband} and topology optimization~\cite{sigmund2003systematic,Bilal_PRE_2011} of the trampoline foundation for different lattice symmetries could be utilized to incur yet additional increases in band-gap size.  
   
This research was supported by the National Science Foundation under Grant No. 0927322 (E. A. Misawa) and Grant No. 1131802 (B. M. Kramer). 


%
%

%


\bibliography{aipsamp5_short}



\end{document}